# Bayesian Inference Applied to the Electromagnetic Inverse Problem

David M. Schmidt,* John S. George and C.C. Wood

Biophysics Group, MS-D454, Los Alamos National Laboratory, Los Alamos, NM 87545

## Abstract

We present a new approach to the electromagnetic inverse problem that explicitly addresses the ambiguity associated with its ill-posed character. Rather than calculating a single "best" solution according to some criterion, our approach produces a large number of likely solutions that both fit the data and any prior information that is used. While the range of the different likely results is representative of the ambiguity in the inverse problem even with prior information present, features that are common across a large number of the different solutions can be identified and are associated with a high degree of probability. This approach is implemented and quantified within the formalism of Bayesian inference which combines prior information with that from measurement in a common framework using a single measure. To demonstrate this approach, a general neural activation model is constructed that includes a variable number of extended regions of activation and can incorporate a great deal of prior information on neural current such as information on location, orientation, strength and spatial smoothness. Taken together, this activation model and the Bayesian inferential approach yield estimates of the probability distributions for the number, location, and extent of active regions. Both simulated MEG data and data from a visual evoked response experiment are used to demonstrate the capabilities of this approach.

## 1. Introduction

Under suitable conditions of spatial and temporal synchronization, neuronal currents are accompanied by electric potentials and magnetic fields that are sufficiently large to be recorded non-invasively from the surface of the head. These are known as the electroencephalogram (EEG) and magnetoencephalogram (MEG), respectively. In contrast to PET and fMRI, which measure cerebral vascular changes secondary to changes in neuronal activity, EEG and MEG are direct physical consequences of neuronal currents and are capable of resolving temporal patterns of neural activity in the millisecond range [Hämäläinen et al., 1993; Aine, 1995; Toga and Mazziotta, 1996]. Unlike PET and fMRI, however, the problem of estimating the current distribution in the brain from surface EEG and MEG measurements (the so-called electromagnetic inverse problem) is mathematically ill-posed; that is, it has no unique solution in the most general, unconstrained case [von Helmholtz, 1853; Nunez, 1981].

Existing approaches to the electromagnetic inverse problem fall into two broad categories: (1) "few-parameter models" (i.e., those in which M ≪ N, where M is the number of parameters to be estimated in the model and N is the number of recording sites); (2) and "many-parameter models" (i.e., those in which M ≥ N). A well-known example of the "few parameter" approach is the single- or multiple-dipole model [e.g., Kavanaugh et al., 1978; Scherg and von Cramon, 1986; Mosher et al., 1992], in which the current is assumed to be represented by a few point-dipoles, the "order" of the model is estimated using Chi-square or related statistical techniques, and the best-fitting values of the dipole parameters (locations, orientations, and magnitudes) are estimated using non-linear numerical minimization techniques. A well-known example of the "many-parameter" approach is the "minimum-norm linear inverse" [e.g., Hämäläinen and Ilmoniemi, 1984; Dale and Sereno, 1993; Hämäläinen et al., 1993], in which the problem is under-determined (because M ≥ N) and a strictly mathematical criterion is used to select among the many solutions that fit the data equally well; in the case of the minimum-norm approach the mathematical criterion is the solution that minimizes the sum of squared current strengths.

In this paper we introduce a new probabilistic approach to the electromagnetic inverse problem, based on Bayesian inference [e.g., Bernardo and Smith, 1994; Gelman et al., 1995]. Unlike other approaches to this problem, including other recent applications of Bayesian methods [Baillet and Garnero, 1997; Phillips et al., 1997], our approach does not result in a single "best" solution to the problem. Rather, we estimate a probability distribution of solutions upon which all subsequent inferences are based. This distribution provides a means of identifying and estimating the likelihood of features of current sources from surface measurements that explicitly emphasizes the multiple solutions that can account for any set of surface EEG/MEG measurements.

---

*To whom correspondence should be addressed, Email: DSchmidt@LANL.GOV

In addition to emphasizing the inherent probabilistic character of the electromagnetic inverse problem, Bayesian methods provide a formal, quantitative means of incorporating additional relevant information, independent of the EEG/MEG measurements themselves, into the resulting probability distribution of inverse solutions. Such information might include constraints derived from anatomy on the likely location and/or orientation of current [Wang et al., 1992; George et al., 1995; Baillet and Garnero 1997; Dale, 1997], maximum current strength, spatial and/or temporal smoothness of current, etc. The Bayesian approach also provides a way to marginalize over nuisance variables that cannot be determined or resolved from the data.

We begin with an overview of the general techniques of Bayesian inference. Then we show how these techniques may be applied to the EEG/MEG inverse problem and demonstrate their use in examples from both simulated MEG data and MEG data from a visual evoked response experiment.

## 2. Bayesian Inference

Bayesian inference (BI) is a general procedure for constructing a (posterior) probability distribution for quantities of interest from the measurements given (prior) probability distributions for all of the uncertain parameters—both those that relate the quantities of interest to the measurements and the quantities of interest themselves. The method is conceptually simple, using basic laws of probability, making its application even to complicated problems relatively straightforward. The posterior probability distribution is often too complicated to be calculated analytically, but can usually be adequately sampled using modern computer techniques, even in problems with many parameters. The method is outlined here, more detailed presentations can be found elsewhere [e.g., Gelman et al., 1995].

The starting point for BI is Bayes' rule of probability:

$$P(\theta, y) = P(\theta \mid y)P(y), \quad (1)$$

where $P(\theta, y)$ is the joint probability distribution for the quantities, $\theta$ and $y$, $P(\theta \mid y)$ is the conditional probability distribution of $\theta$ given $y$, and $P(y)$ is the marginalized probability distribution of y; $P(y) = \sum_\theta P(\theta, y)$ (or $P(y) = \int P(\theta, y) \, d\theta$ for continuous $\theta$). If $\theta$ represents parameters about which we wish to learn and $y$ represents data bearing upon $\theta$, then the probability of $\theta$ given $y$ can be constructed from Bayes' rule as:

$$P(\theta \mid y) = \frac{P(\theta, y)}{P(y)} = \frac{P(y \mid \theta)P(\theta)}{P(y)}. \quad (2)$$

Here $P(\theta)$ is the prior probability distribution of $\theta$ which represents one's knowledge of $\theta$ prior to the measurement. This is modified by the data through the likelihood function, $P(y \mid \theta)$, to produce the posterior probability distribution, $P(\theta \mid y)$. Since $P(y)$ is independent of $\theta$ it can be considered a normalizing constant and can be omitted from the unnormalized posterior density:

$$P(\theta \mid y) \propto P(y \mid \theta)P(\theta). \quad (3)$$

As summarized in [Gelman et al., 1995], "These simple expressions encapsulate the technical core of Bayesian inference: the primary task of any specific application is to develop the model $P(\theta, y)$ and perform the necessary computations to summarize $P(\theta \mid y)$ in appropriate ways."

## 3. Bayesian Inference Applied to the EEG/MEG Inverse Problem

### 3.1. Activity Model

The methods of BI applied to the EEG/MEG inverse problem are demonstrated within the context of a model for the regions of activation which is intended to be applicable in evoked response experiments. There is both theoretical and experimental evidence that EEG and MEG recorded outside the head arise primarily from neocortex, in particular from apical dendrites of pyramidal cells [e.g., Allison et al., 1986; Dale and Sereno, 1993; Hämäläinen et al., 1993]. We therefore construct a model that assumes a variable number of variable size cortical regions of stimulus-correlated activity in which current may be present. Specifically, an active region is assumed to consist of those locations which are identified as being part of cortex and are located within a sphere of some radius $r$ centered on some location $w$, also in cortex. There can be any number $n$ of these active regions up to some maximum $n_{max}$, and the radius can have any value up to some maximum, $r_{max}$. The goal is to determine the posterior probability values for the set of activity parameters $\boldsymbol{\alpha} = \{n, w, r\}$ which govern the number, location, and extent of active regions.

### 3.2. Probability Model for Activity Parameters

The first step in BI is to construct a probability model that relates the activity parameters to the measurements. Let the $N$ measurements at one instant in time be denoted by $\mathbf{b} = \{b_1, \ldots, b_N\}$. The conditional probability of the activity parameters given the observed data, $P(\boldsymbol{\alpha} \mid \mathbf{b})$, can be expressed using Bayes' rule of probability as

$$P(\boldsymbol{\alpha} \mid \mathbf{b}) \propto P(\mathbf{b} \mid \boldsymbol{\alpha})P(\boldsymbol{\alpha}) \quad (4)$$

where $P(\boldsymbol{\alpha})$ is the prior probability for the activity parameters and $P(\mathbf{b} \mid \boldsymbol{\alpha})$ is the probability of the data given a particular set of values for the activity parameters. The prior probability for the activity parameters will be set by the experimenter using physiological information about the particular experiment being analyzed. Because the data do not depend on the activity parameters directly, but rather on a given current distribution $\mathbf{j}$, the function, $P(\mathbf{b} \mid \boldsymbol{\alpha})$, can not be specified until first expanding it to include the dependence of the measurements on the current. This may be accomplished by marginalizing out the current in the joint probability of the data and current such that

$$
\begin{aligned}
P(\mathbf{b} \mid \boldsymbol{\alpha}) &= \int P(\mathbf{b}, \mathbf{j} \mid \boldsymbol{\alpha}) \, \mathcal{D}\mathbf{j} \\
&= \int P(\mathbf{b} \mid \mathbf{j}, \boldsymbol{\alpha}) P(\mathbf{j} \mid \boldsymbol{\alpha}) \, \mathcal{D}\mathbf{j} \quad (5)
\end{aligned}
$$

where the integral is a functional integral over all current distributions. The function, $P(\mathbf{b} \mid \mathbf{j}, \boldsymbol{\alpha})$, is the likelihood function of the data; there is no explicit dependence upon $\boldsymbol{\alpha}$ since $\mathbf{j}$ is all that is needed to completely specify $P(\mathbf{b} \mid \mathbf{j}, \boldsymbol{\alpha})$. In particular, since most evoked response experiments are repeated many times and averaged, it is assumed that the likelihood function is Gaussian such that

$$
P(\mathbf{b} \mid \mathbf{j}, \boldsymbol{\alpha}) \propto
$$
$$
\exp\left\{ -\frac{1}{2} \sum_{k,l=1}^{N} (b_k - \langle \mathbf{a}_k, \mathbf{j} \rangle) \, C_{kl}^{-1} (b_l - \langle \mathbf{a}_l, \mathbf{j} \rangle) \right\}. \quad (6)
$$

Here, $\mathbf{C}$ is the covariance matrix of the noise or background in the measurements and $\mathbf{a}$ are the forward fields or measurement kernel such that if there were no noise or background the measurements would be related to the current by the inner product:

$$
b_k = \langle \mathbf{a}_k, \mathbf{j} \rangle = \int \mathbf{a}_k(\mathbf{x}) \cdot \mathbf{j}(\mathbf{x}) \, d^3x. \quad (7)
$$

We find it convenient to use an equivalent representation of Eq. 6 which has the noise covariance absorbed into a new set of effective measurements and forward fields, $\tilde{\mathbf{b}}$ and $\tilde{\mathbf{a}}$, such that Eq. 6 becomes

$$
P(\mathbf{b} \mid \mathbf{j}, \boldsymbol{\alpha}) \propto \exp\left\{ -\frac{1}{2} \sum_{k=1}^{N} \left( \tilde{b}_k - \langle \tilde{\mathbf{a}}_k, \mathbf{j} \rangle \right)^2 \right\}. \quad (8)
$$

The function $P(\mathbf{j} \mid \boldsymbol{\alpha})$, which gives the probability of any current given a particular set of activity parameters, needs to be constructed. Clearly the current should be zero outside of active regions. Furthermore, we would like to be able to incorporate prior information about the limits of current strength, spatial variability and orientation of the current within active regions. For example, high-resolution current source density estimates suggest that net cortical currents are oriented predominantly perpendicular to the cortical surface [Mitzdorf, 1985]. These forms of prior information may be incorporated, in a manner which simplifies computing the marginalization over $\mathbf{j}$ in $P(\mathbf{b}, \mathbf{j} \mid \boldsymbol{\alpha})$, by using a Gaussian distribution such that

$$
P(\mathbf{j} \mid \boldsymbol{\alpha}) \propto |\mathbf{V}_{\boldsymbol{\alpha}}|^{-\frac{1}{2}} \exp\left\{ -\frac{1}{2} \langle \mathbf{j}, \mathbf{V}_{\boldsymbol{\alpha}}^{-1} \mathbf{j} \rangle \right\} \quad (9)
$$

where $\mathbf{V}_{\boldsymbol{\alpha}}^{-1}$ is the inverse of the covariance operator (matrix) of the current. The diagonal elements, or the variances, serve to limit the current strength, and the off-diagonal elements, which are related to the correlation coefficients, can serve to restrict the smoothness and orientation of the current distribution. The variance at locations which are not part of any active region for a given $\boldsymbol{\alpha}$ is set to zero. The experimenter needs to set the values of the covariance matrix, based on knowledge of the the experiment to be analyzed, using prior information about the strength, orientation and spatial variability of current within active regions.

The full probability model for the activity parameters is

$$
P(\boldsymbol{\alpha} \mid \mathbf{b}) \propto P(\boldsymbol{\alpha}) |\mathbf{V}_{\boldsymbol{\alpha}}|^{-\frac{1}{2}} \times
$$
$$
\int \exp\left[ -\frac{1}{2} \left\{ \sum_{k=1}^{N} \left( \tilde{b}_k - \langle \tilde{\mathbf{a}}_k, \mathbf{j} \rangle \right)^2 + \langle \mathbf{j}, \mathbf{V}_{\boldsymbol{\alpha}}^{-1} \mathbf{j} \rangle \right\} \right] \mathcal{D}\mathbf{j} \quad (10)
$$

and $P(\boldsymbol{\alpha})$ is set by the experimenter. The integral over the current may be constructed using the eigenvalues, $\{\lambda_\theta(\boldsymbol{\alpha})\}$, and normalized eigenvectors, $\{\boldsymbol{\psi}_\theta(\boldsymbol{\alpha})\}$ $(\theta = 1, \dots, N)$, of the matrix $G_{k,l}(\boldsymbol{\alpha}) = \langle \tilde{\mathbf{a}}_k, \mathbf{V}_{\boldsymbol{\alpha}} \, \tilde{\mathbf{a}}_l \rangle$; all of which may be calculated using standard numerical techniques. Using these eigenvalues and eigenvectors the formula for the posterior probability distribution becomes

$$
P(\boldsymbol{\alpha} \mid \mathbf{b}) \propto P(\boldsymbol{\alpha}) \times
$$
$$
\exp\left[ -\frac{1}{2} \left\{ \sum_{k,\theta,l} \tilde{b}_k \frac{\psi_{k,\theta}(\boldsymbol{\alpha}) \psi_{l,\theta}(\boldsymbol{\alpha})}{1 + \lambda_\theta(\boldsymbol{\alpha})} \tilde{b}_l + \sum_\theta \ln(1 + \lambda_\theta(\boldsymbol{\alpha})) \right\} \right]. \quad (11)
$$

This formula is well-behaved and is not overly sensitive to very small eigenvalues. Moreover, it is relatively simple to compute because it only depends on the $N$ by $N$ matrix, $G_{k,l}(\boldsymbol{\alpha})$.

### 3.3. Sampling the Posterior

The next step in BI is to use the posterior probability distribution in order to answer questions related to the activity parameters in terms of probability. Examples of such questions include: what is the probability that there were $m$ regions of activity? What are the locations for these active regions at a 95% probability level? In cases where the number of different possible sets of activation parameters is small, one can evaluate the complete posterior distribution. Generally, however, the number of different possible sets of activation parameters is large. In such cases the method of Markov Chain Monte Carlo (MCMC) can be used to generate a sample of sets of activity parameters which are distributed according to the posterior distribution. This is known as sampling the posterior, the techniques for which are described in detail elsewhere [e.g., Gelman et al., 1995].

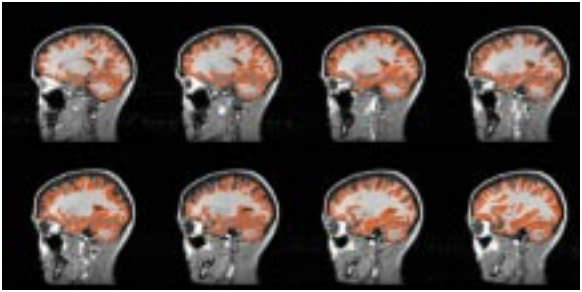

Fig. 1: Gray matter regions are tagged (in red) from anatomical MRI data. These tagged voxels constitute the anatomical model used to implement the cortical location and orientation prior information.

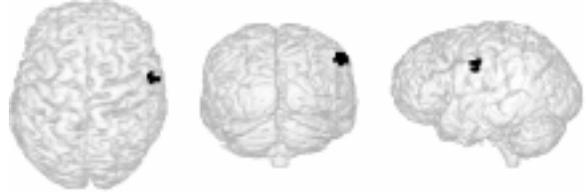

Fig. 2: Maximum intensity projection of the location and extent (in black) of the active region used to generate simulated MEG data for Example 1.

## 4. Examples

While the methods just described apply to models for both EEG and MEG data, in the remainder of this paper we will use MEG data to illustrate the properties of the approach. Both simulated and empirical MEG data for a Neuromag-122 whole-head system were used. The physical setup of the actual MEG experiment was used to determine the location of the subject's head relative to the sensors in the simulated data examples. In addition, an anatomical MRI data set acquired from the subject in the MEG experiment was used to determine the location of cortex (actually gray matter) using MRIVIEW (Fig 1), a software tool developed in our laboratory [Ranken and George, 1993]. About 50,000 voxels were tagged and the normal directions for each of these voxels was then determined by examining the curvature of the local tagged region.

A spherically symmetric conductivity model was used to calculate the expected measurements given a current source both for the simulated data and in the likelihood calculations [Sarvas, 1987]. The same prior assumptions were used with the simulated data sets, with only minor changes for the real data example. Specifically the prior probability function $P(\boldsymbol{\alpha})$ was uniform so that each set of activation parameters had the same prior probability. The number of active regions was allowed to range from 0 to 8 and the radius of any region of activity was allowed to range from 0 to 10 mm.

The covariance matrix was factored such that

$$^{\beta\gamma}V_{\boldsymbol{\alpha}}(i,j) = \sigma_{\boldsymbol{\alpha}}(i)\sigma_{\boldsymbol{\alpha}}(j)\rho(i-j)\,^{\beta\gamma}\Omega(i,j) \quad (12)$$

where $\beta$ and $\gamma$ are orientation indices, $i$ and $j$ are location indices, $\sigma_{\boldsymbol{\alpha}}(i)$ is the standard deviation at location $i$, $\rho(i-j)$ is the spatial correlation function, and $^{\beta\gamma}\Omega(i,j)$ is the orientation covariance. The correlation function was chosen to be a Gaussian with zero mean and 7 mm standard deviation which imposes spatial smoothness on scales of about 7 mm or less. Because of this prior information concerning spatial correlation, the continuous current distributions and integrals of the previous section may be well-approximated by discrete distributions and sums over the volume elements (voxels) that were tagged from the anatomical MRI data. For example, in evaluating the posterior probability value using Eq. 11 the matrix $\mathbf{G}$ is calculated in the following examples by approximating the continuous integral with a sum over tagged voxels. This is a good approximation because the covariance operator has a correlation length of 7 mm which is larger than the voxel dimensions of 2 mm on a side.

To complete the specification of the covariance operator, a value of 2 nAm was used for $\sigma_{\boldsymbol{\alpha}}(i)$ at all locations in active regions and 0 nAm elsewhere. The orientation covariance was chosen such that there was no correlation between the orientations at different locations and the orientation distribution at any given location was symmetric with respect to the direction normal to the cortical surface at that location and had a mean equal to the cortical norm direction and a standard deviation of 30°. Unlike other recent implementations of cortical constraints in distributed inverse solutions [e.g., Dale and Sereno, 1993; Baillet and Garnero, 1997], this procedure results in a distribution of orientations around the perpendicular, not a fixed normal orientation.

Finally, the same noise was added to all simulated data sets, which was Normal with a standard deviation of 10 fT. The values used here in the prior probability distribution are meant to be an example of what one might choose for a MEG analysis and should be chosen for each particular MEG data set.

### 4.1. Example 1

The location and extent of the active region used to generate the simulated MEG data is shown in Fig. 2. The bounding radius of the active region was 5 mm and the current dipole strength at each voxel was 2 nAm oriented in the cortical normal direction. A plot of the simulated data and its field pattern is shown in Fig. 3. Ten thousand samples were drawn from the posterior distribution using a MCMC algorithm. A plot of the posterior probability for each of the samples is shown in Fig. 4. It took about 600 samples to progress from the starting point which had a low probability to one that had a high probability and was therefore representative of the posterior distribution. Only the final 9,000 samples, a few of which are shown in Fig. 5, were used in making probabilistic inferences as discussed below. All of the samples shown in Fig. 5 are among the 95% most probable and therefore fit both the data and the prior expectations quite well. Any of these could have produced the given MEG data, yet there are clearly vast differences among the samples. The number of active regions ranges from 1 to 5, the sizes of the regions vary greatly and the locations of the active regions vary nearly across the entire tagged region of the brain (when considering all 9,000 samples). This variability is a representation of the degree of the ambiguity of the inverse problem for these MEG data, even with the prior information present.

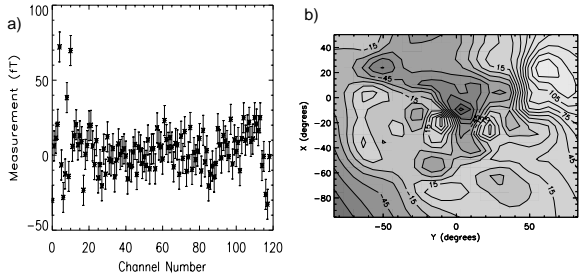

Fig. 3: The simulated data used in Example 1, a) as a function of channel number, b) as a field pattern as viewed from a top projection.

Despite the degree of variability among the samples in Fig. 5 a property common to all is apparent; namely an active region in the dorsal, lateral region of the right hemisphere. A feature, such as this, common to all or most of the samples, is associated with a high degree of probability. This probability can be quantified because the MCMC samples are distributed according to the posterior probability distribution. The smallest set of voxels which contains the center of the active region in the dorsal, lateral region in 95% of the samples was identified and is shown in Fig. 6. This region, which contains a center of activity with a probability of 95%, in fact encompasses the region of activity which was used to produce the simulated data set (Fig. 2). Although it is nice to see this agreement, it is not sufficient to justify this or any MEG inverse method based solely on whether it produces results consistent with the true active re-gions because any of the sets of active regions shown in Fig. 5 could have also been used to generate the same MEG data. Any robust and highly probable result or inference therefore should be consistent with the wide range of possible sets of active regions, as is the result in Fig. 6 by construction. This is a very important feature of BI which is necessarily missing from any other analysis method which only considers just one possible result, even if it happens to be the most likely result within a given model.

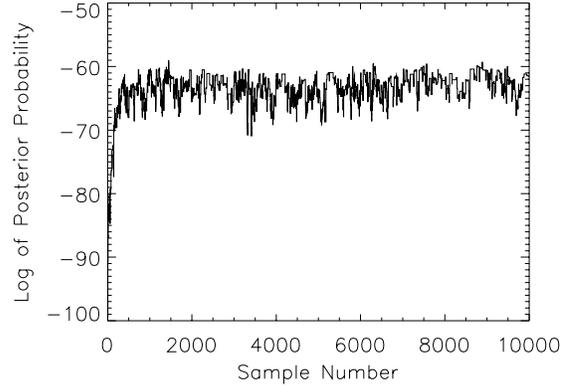

Fig. 4: The posterior probability of the 10,000 samples drawn from the posterior probability distribution of Example 1. This figure shows the progression of the MCMC sampling algorithm.

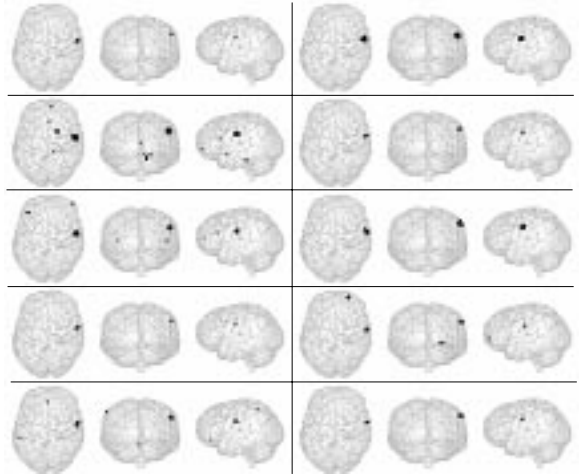

Fig. 5: A few of the 9,000 samples drawn from the posterior probability distribution of Example 1. Each panel shows 3 views of the maximum intensity projection of active regions from a single sample. All of these samples could have produced the same MEG data set.

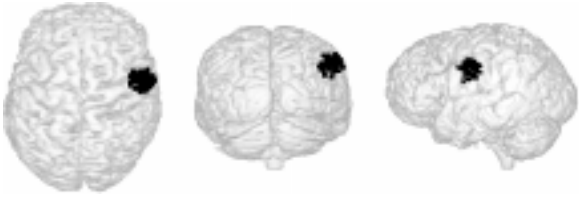

Fig. 6: Maximum intensity projections of the location and extent of a region containing a center of activity at a 95% probability level in Example 1.

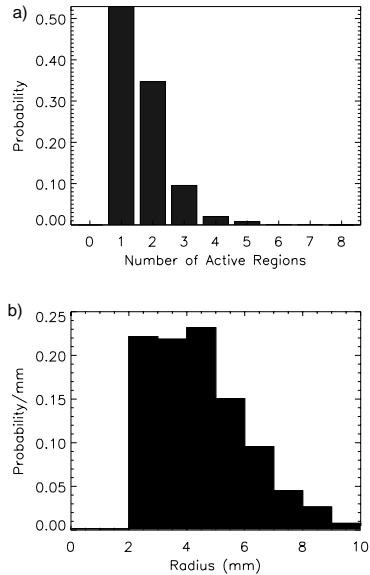

Fig. 7: The posterior probability for a) the number of active regions present in Example 1 and b) the radius of the sphere bounding activity in Example 1, assuming there was only one active region present..

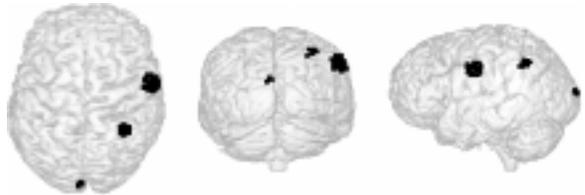

Fig. 8: Maximum intensity projections of the location and extent of the active regions used to generate the simulated MEG data for Example 2.

In addition to the information about the locations of probable regions of activity, the Bayesian approach combined with this activity model also provides probabilistic information about the number and size of active regions. The posterior distribution for the number of active regions was constructed by histograming the number of regions across the MCMC samples. This histogram is shown in Fig. 7a. One active region is the most probable; however, two active regions are quite likely as well. Although the location of one active region was identified in the last paragraph, the location of a second could not be well localized because it occurred in a wide range of locations across the MCMC samples. Assuming there was only one active region present, we can obtain information about its size by histograming the size of active regions in the MCMC samples that had only one region present. This histogram is shown in Fig. 7b and represents the posterior probability for the bounding radius of the active region, assuming that there was only one region active. Regions smaller than 2 mm and larger than 8 mm in radius are very unlikely whereas regions that are around 5 mm in radius are likely. The size of the region used to produce the simulated data was 5 mm. We believe that much of the information on size derives from prior information about location, orientation and strength of neural current.

Other inferences could be drawn using the MCMC samples in a similar manner. For example, one could construct the probability for the size of the active region, assuming there was one centered within the 95% probability region shown in Fig. 6, rather than assuming there was only one active region present throughout the entire head as was done above.

### 4.2. Example 2

A second simulated data set was generated using the three active regions of different sizes shown in Fig. 8. The most anterior region is centered at the same location as the region in the first example except for this case it has a bounding radius of 8 mm. A current dipole strength of 2 nAm oriented normal to cortex was used at each voxel within this bounding sphere. The nearby, more posterior region had a bounding radius of 5 mm and a current dipole strength of 2.5 nAm was used. The most posterior region had a 3 mm bounding radius and a current dipole strength of 1.5 nAm. The same noise and prior assumptions where used here as for the first example. Figure 9 shows a plot of the resulting simulated data and the field pattern.

Ten thousand samples were drawn from the posterior of which the final 8,000 were used to make probabilistic inferences. Just as in the first example we expect there to be many different locations where activity may be found in these samples. Since we are interested in those locations which contain activity in most of the samples it is useful to make a histogram of the locations of the centers of active regions across the 8,000 samples. This histogram is shown in Fig. 10. It is relatively simple to determine those re-gions that contain centers of activity at a 95% probability level from this histogram by centering each region on the local peaks in the histogram and expanding the radius of each region until a 95% level is reached. This was done for the 3 peaks present in Fig. 10 and is shown in Fig. 11. While these regions are consistent with the locations of the active regions used to generate the simulated data, what is more important is that these regions are consistent with at least 95% of the likely sets of active regions that could have also generated this data. This is true even when allowing a variable number of active regions of variable extent. Furthermore, these regions are not necessarily the only regions that could have been active. As shown in Fig. 12 there is significant probability that more than 3 regions may have been active. What is shown in Fig. 11 are the locations of those active regions that occurred consistently in well localized areas across the MCMC samples. Other possible active regions were not so well localized.

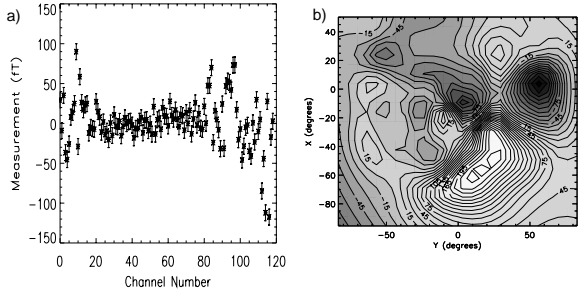

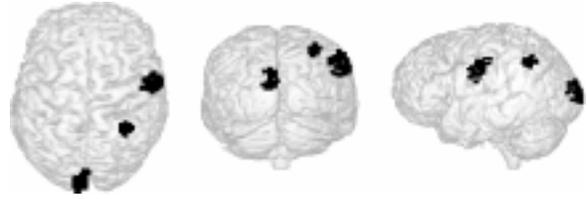

Fig. 11: Maximum intensity projections of the location and extent of the three regions that were found to contain centers of activity at a probability level of at least 95% in Example 2.

Fig. 9: The simulated data of Example 2, a) as a function of channel number and b) as a field pattern in a top projection view.

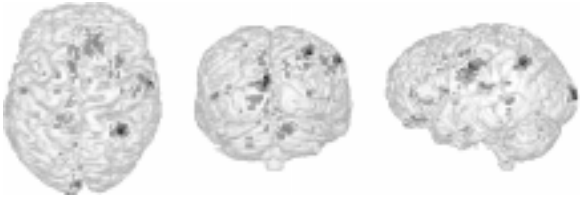

Fig. 10: Maximum intensity projections of the histogram of centers of active regions across the MCMC samples in Example 2, shown on top of surface renderings of cortex. The darker the shade of a region the larger the value of the histogram at that location.

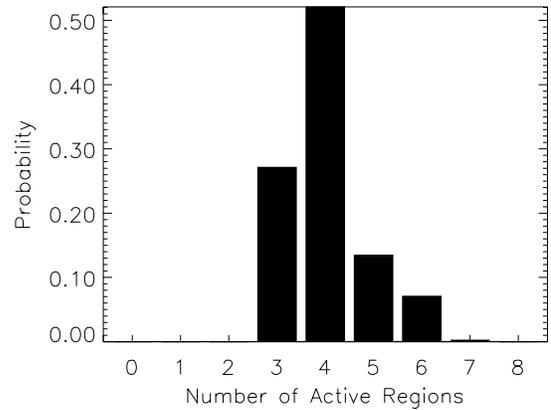

Fig. 12: The posterior probability for the number of active regions in Example 2.

In order to learn about the extent or size of each of the active regions localized in Fig. 11 a histogram of the radius of the active regions present in each of the areas shown in Fig. 11 across the samples was made. This represents the posterior probability for the size of active regions, assuming there was an ac-tive region in each of these areas. These plots are shown in Fig. 13. Recall that the radii of the actual regions used to generated the data were 8 mm, 5 mm and 3 mm for the regions in anterior to posterior order. The agreement between actual radii and posterior probabilities is especially remarkable given the variation in the current strengths of the regions used to generate the data. Such information on extent can be very useful, is not present in most other current methods for analyzing MEG data, and is affirmation of the likely utility of anatomical and physiological prior information.

### 4.3. Example 3

The final example, which is based on MEG data from a visual evoked response experiment [Aine et al., 1997a], illustrates the feasibility and the value of the approach with actual data. In order to examine the sensitivity of the Bayesian approach to detect known features of human visual cortex organization, we compared Bayesian analyses of MEG responses to visual stimuli in the left and right visual fields. Based on the crossed anatomical projections of the visual fields to the brain and on previous lesion, MEG, and fMRI studies in humans [e.g., Horton and Hoyt, 1991; Sereno et al., 1995; Aine et al., 1996], initial cortical activation for stimuli in the left and right visual fields should occur near the calcarine fissure in the occipital region of the contralateral hemispheres.

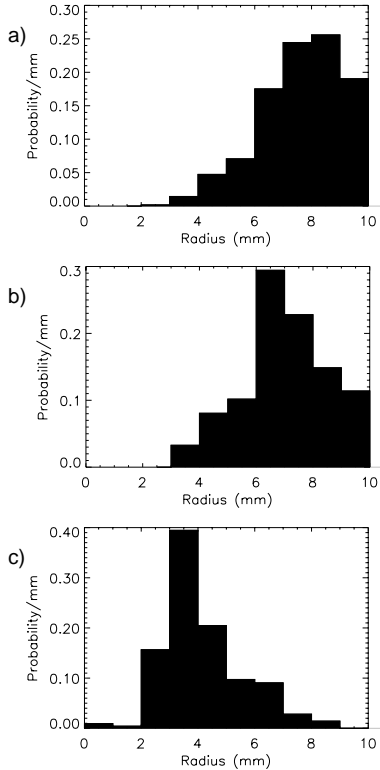

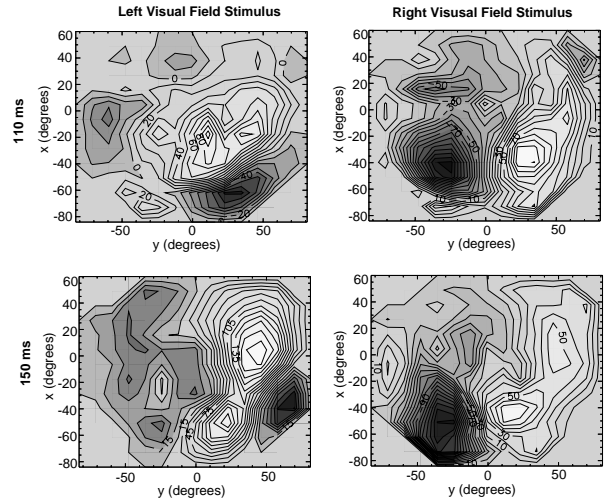

Fig. 14: The field patterns for the real MEG data in Example 3. The iso-amplitude contours are surface projections on a plane; $y = 0$ represents the mid-line on the top of the head. Positive fields (flux emerging from the head) are displayed in light shades and negative fields (reentering flux) are shown in dark shades.

Fig. 13: The posterior probability distributions for the size of the three active regions whose centers are shown in Fig. 11 in anterior to posterior order. The true sizes of the regions used to generate the simulated data were a) 8 mm, b) 5 mm and c) 3 mm, respectively.

The visual stimuli were black-white circular sinusoidal patterns, $1.0°$ in diameter, presented near the horizontal meridian at $6.2°$ in the left and right visual fields. The stimulus duration was 250 ms and the average inter-stimulus interval was 1.0 s. One hundred epochs (from 100 ms before each stimulus to 400 ms after each stimulus) were averaged; bad channels were identified and removed before data analysis. The variance of the noise was estimated by calculating the variance of the pre-stimulus epoch. The same model and the same prior information used in Examples 1 and 2 were used for the Bayesian analyses in this example, except that the standard deviation of the current strength was assumed to be 8 nAm instead of 2 nAm. This value is is consistent with the maximum current strength measurements in [Okada

et al., 1998].

The model was applied separately to the data for each visual field stimulus at 10 ms intervals from 110 ms to 160 ms post-stimulus. Ten thousand samples of the posterior probability were generated for each latency. The results to be presented here are from 110 ms and 150 ms following stimulus onset; latencies that should include robust activation of the calcarine region [Aine et al., 1996]. Figure 14 presents the field distributions for these data.

The top of Fig. 15 presents maximum intensity projections of the probability of activity for each voxel in the anatomical model for the left and right visual field stimuli at two different latencies following stimulus onset (110 and 150 ms, respectively). This probability distribution was constructed by calculating the fraction of MCMC samples in which each voxel had activity and is a marginalization of the full posterior probability distribution onto the space of anatomical voxels. The bottom of Fig. 15 presents the posterior probability marginalized onto the number of active regions for each latency and visual field combination.

For the left visual field stimulus, maximal probability of activation at 110 ms was located in the right (contralateral) hemisphere, centered upon the calcarine region. This pattern was reversed for the right visual field stimulus at 110 ms, consistent with the predictions from anatomy, and from the lesion, fMRI, and previous MEG studies cited above. In order to show this more clearly, regions which contained activity at a probability level of 95% were identified and are shown in greater detail in Fig. 16, which depicts relative probability of activation within these regions on a color scale in three orthogonal slices through the calcarine region and a three-dimensional rendering of the occipital region.

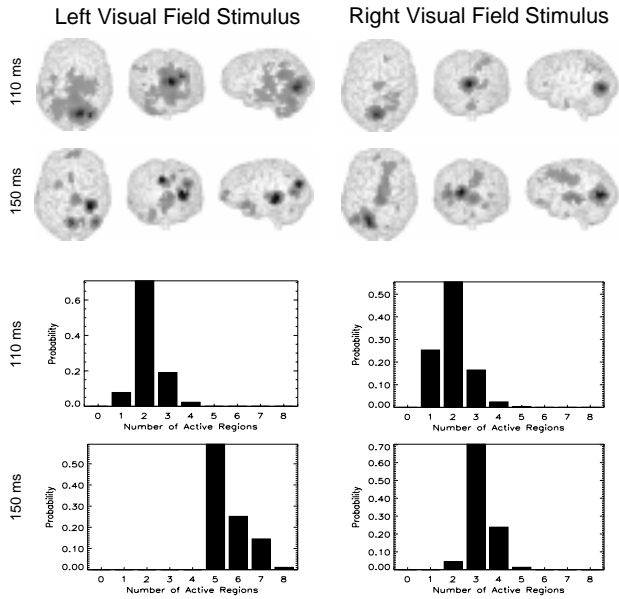

Fig. 15: Posterior probability distributions for the real MEG data of Example 3, marginalized onto anatomical location and onto number of active regions. The probability of activation as a function of location distributions are shown as maximum intensity projections over surface renderings of anatomy in the top half of the figure. Here, darker shades of grey indicate higher probability. The distributions for the number of active regions are shown in the bottom half of the figure. These results show evidence for activation contralateral to the stimulus at both 110 ms and 150 ms latencies.

For both the left and right visual field stimuli, the most probable number of active regions at 110 ms latency was two, suggesting that active regions in addition to the most probable ones in the calcarine regions of each hemisphere were needed to account for the data and prior information. However, these regions were inconsistently located over the Monte Carlo samples, as indicated by the relatively widespread regions of low probability in addition to the focus of high probability in Fig. 15.

At 150 ms, the most probable number of active regions increased to five for the left visual field stimulus and to three for the right visual field stimulus. Regions of highest probability in each case were located in parieto-occipital and temporo-occipital regions of the hemisphere contralateral to the visual field stimulated. These results are consistent in general terms with MEG and fMRI evidence of multiple regions of extra-striate activity [Aine et al., 1997a; Aine et al., 1997b; Shah et al., 1998], although much additional work is needed to obtain a definitive comparison of Bayesian inference, multiple-dipole, and fMRI estimates of activity in such experiments.

Two additional features of the results in Example 3 should be noted. First, although maximal probability of activation at the 110 ms latency was indeed located in the opposite hemisphere, there exists sizable probability for activity in the ipsilateral hemisphere near the mid-line. The extent of the 95% probability regions shown in Fig. 16 is indicative of both the extent of estimated activation and the degree of error or uncertainty in that estimate even allowing for the possibility of different numbers of active regions of variable extent. Second, although not shown in detail here, analyses at other latencies suggest a progressively increasing number of probable regions of activation, in both the ipsilateral and contralateral hemispheres, over the latency region from 110 to 160 ms following stimulus onset. It will be of considerable interest to explore the time dependence of the Bayesian inference analyses in relation to evidence for multiple, functionally organized areas of striate and extra-striate visual cortex and to examine the value of temporal prior information (not included in the current activation model) in the form of, for example, temporal covariance constraints.

## 5. Discussion

We have demonstrated a method for analyzing EEG/MEG data that directly addresses the ill-posed character of the electromagnetic inverse problem by allowing probabilistic inferences to be drawn about regions of activation from a large number of possible solutions which both fit the data and the prior expectations made explicit by the Bayesian approach. In addition, we have introduced a model for the current distributions corresponding to neural activity that produce MEG (and EEG) data that is not overly restrictive, allows extended regions of activity, and can easily incorporate prior information such as anatomical constraints from MRI [Dale and Sereno, 1993; George et al., 1995].

Other investigators have applied the Bayesian formalism to various models for EEG or MEG inverse problems [e.g., Phillips et al., 1997; Baillet and Garnero, 1997]. Any Bayesian approach requires: (a) a model that relates the EEG/MEG measurements to underlying neuronal currents; and (b) an implementation of that model within the Bayesian formalism, including the nature and parameterization of independent prior information. The approach presented here differs from those of [Phillips et al., 1997] and [Baillet and Garnero, 1997] both in the form of the activity model employed and in the manner in which the Bayesian formalism is exploited. Much additional work is needed to determine which combination of source model and prior information is most useful for the electromagnetic inverse problem. However, we believe it is clear even at this early stage that the strategy of estimating the probability distributions for model parameters (e.g., the number, location, and extent of active regions in the present model) is a richer, more robust, and more realistic basis for inference than estimating a single, "best-fitting" solution within a given model. As we have demonstrated, considering only one solution, even if it is the most likely, is not necessarily representative of the range of possible solutions that both fit the data and the prior information. Only by considering this full range of possible solutions can one construct robust, reliable inferences from the data.

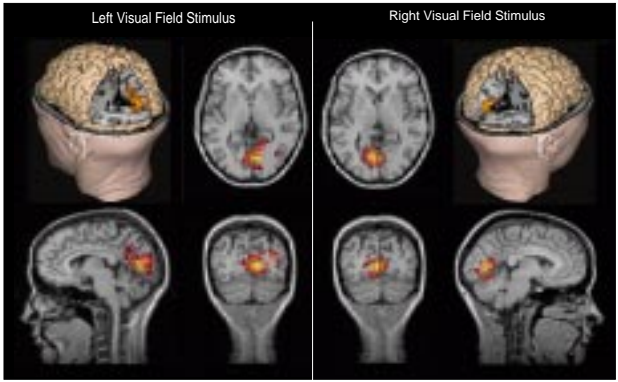

Fig. 16: Four views of a region that was found to contain activity at a 95% probability level in Example 3, for both a left and a right visual field stimulus, at 110 ms latency. The two-dimensional views show the regions (in color) within the anatomical MRI data (grey-scale). Shades of color represent relative probability within the regions on a temperature-like scale; bright yellow represents the highest probability. The horizontal and coronal views are from the top and from the back of the subject, respectively; the sagital views are from the left for the left visual field stimulus panel and from the right for the right visual field stimulus panel. The three-dimensional views are useful for showing the location of the regions relative to other brain structures. These results indicate that the probability of activity is maximal in the calcarine region of the hemisphere contralateral to the visual field stimulated.

Bayesian approaches to data analysis in general are often criticized for the lack of objectivity associated with prior information [e.g., Efron, 1986]. Those criticisms apply as well to Bayesian applications of the electromagnetic inverse problem and it is essential to attempt to justify both the choice of the source model and the nature and specific values of the prior information as thoroughly and rigorously as possible. However, it is important to note that any attempt to solve the electromagnetic inverse problem forces the investigator to make analogous assumptions, even though they are rarely explicit. For example, widely used inverse approaches such as dipole models, minimum norm, FOCUSS [Gorodnitsky et al., 1995] or LORETA [Pascual-Marqui et al., 1994] all require restrictive assumptions regarding the nature and form of the allowable current distributions. A Bayesian approach: (a) generalizes this strategy by weighting the possible current distributions on a probabilistic continuum instead of restricting the possibilities to those that are allowed; and (b) requires that the assumptions and prior information be made explicit and their associated prior probability distributions be justified explicitly. This formal, explicit, treatment of prior information in Bayesian approaches is therefore a useful general feature for applications to inverse problems.

Finally, we emphasize that the activity model and examples described here are meant to illustrate the techniques and capabilities of BI in EEG/MEG and that other activity models, conductivity models or sets of parameters of interest might be more appropriate for different experimental conditions. Our major objective in this paper has been to present and illustrate the value of the Bayesian inferential approach, not to argue for the universal applicability of the particular activity model and prior information employed. Nevertheless, we believe the activity model described here is useful for many functional imaging applications and can readily be extended in a number of ways. These include incorporating temporal prior information in the form of temporal covariance constraints [e.g., Dale and Sereno, 1993] or explicit temporal models for evoked response studies. In addition, the Bayesian approach provides a natural means for incorporating information from other functional imaging modalities such as PET or fMRI [George et al., 1995; Belliveau, 1997; Dale, 1997]. The latter can be readily achieved with the Bayesian framework and with this activity model by assigning prior probabilities to possible locations of active regions based on results from the other modality or modalities. Such a Bayesian formulation of multimodality integration would yield an inherently probabilistic result in which the quantity estimated would be the probability of activation as a function of both space and time.

## Acknowledgments

We thank Cheryl Aine for use of the visual evoked response MEG data, Elaine Best for help with preprocessing this data, and Doug Ranken for help with some of the figures. Aspects of this work have been presented at the Tenth International Conference on Biomagnetism (Biomag96) and at the second and third international conferences on Functional Mapping of the Human Brain. Supported by: NIH Grants EY0861003 and DA/MH09972, the Los Alamos National Laboratory and the United States Department of Energy.